\documentclass[aps,pra,showpacs,reprint,superscriptaddress,floatfix]{revtex4-1}
\usepackage{graphicx}
\usepackage{longtable}
\usepackage{dcolumn}
\usepackage{multirow}
\usepackage{xcolor}
\usepackage{amssymb, amsmath}
\usepackage[colorlinks=true, citecolor=blue,linkcolor=blue]{hyperref}
\begin{document}
\title{Calculations of $\mathcal P$ and $\mathcal T$\,-odd interaction constants of alkaline\,-\,earth monofluorides using KRCI method }
\author{Renu Bala}
\email{rbala@ph.iitr.ac.in}
\author{H. S. Nataraj}
\email{hnrajfph@iitr.ac.in}
\affiliation{ Department of Physics, Indian Institute of Technology Roorkee,
Roorkee - 247667, India}
\author{Malaya K. Nayak}
\email{ mknayak@barc.gov.in, mk.nayak72@gmail.com}
\affiliation{Theoretical Chemistry Section, Chemistry Group, Bhabha Atomic Research Centre, Trombay Mumbai 400085, India}
\begin{abstract}
We have reported the results of \textit{ab initio} calculations of parity- and time\,-\,reversal\,-odd interaction constants for the ground state of alkaline\,-\,earth monofluorides. The Kramers\,-\,restricted configuration interaction method limited to single and double excitations in conjunction with the quadruple zeta quality basis sets have been employed to perform these $4$\,-\,component relativistic calculations. The results are compared with the existing semi\,-\,empirical and other theoretical results, wherever available. \\

\end{abstract}
\maketitle
%
%-------------------------------------------%
\section{Introduction}
%-------------------------------------------%
%
The parity- and time\,-\,reversal ($\mathcal P$ \& $\mathcal T$) symmetry violating effects such as the electric dipole moment of an electron (eEDM) and the scalar\,-\,pseudoscalar ($S$\,-\,$PS$) interactions between nucleons and electrons which manifest in giving rise to the intrinsic electric dipole moment (EDM) of atoms and molecules. The effects will be more pronounced in heavier atomic/molecular systems. Numerous atoms and molecules have thoroughly been scrutinized for the observation of such effects~\cite{Bernreuther, Sunaga1, Meyer1, Chupp, Baron, Hudson, Cairncross, Hudson1, DeMille1, Altuntas} as they hold answers to some of the fundamental mysteries of our Universe~\cite{DeMille, Fuyuto, Chupp}.
%%%%%%%%%%%%%%%%%%%%%

In spite of more than seven decades of laborious efforts by several experimental groups, the conclusive evidence of a non-zero eEDM has mostly been elusive. In this context, the heavy open\,-\,shell polar molecules would be considered more suitable than atoms because of the large values of effective intrinsic electric fields ($\varepsilon_{eff}$)~\cite{Sandars} in the former. The high precision measurements carried out using diatomic molecules such as ThO, HfF$^+$ and YbF molecules~\cite{Baron, Cairncross, Hudson1} have yielded best limits  on eEDM so far. 
An accurate knowledge of $\mathcal{P}$ \& $\mathcal{T}$\,-\,odd interaction constants such as: $W_d$ that characterizes $\varepsilon_{eff}$ and $W_s$ that characterizes the $S$\,-\,$PS$ interaction between  the nucleons and electrons, is required to interpret the results of the experiments. Invocation of accurate quantum chemical methods will be necessary for the calculation of such interaction constants.
%%%%%%%%%%%%%%%%%%%%%

Among several diatomic molecules that have been considered, the alkaline\,-\,earth monofluorides (AEMFs) have a special place since laser cooling and trapping experiments have been  or being performed for MgF~\cite{Yin}, CaF~\cite{Zhelyazkova, Tarbutt}, SrF~\cite{Barry, Barry1, Shuman, Truppe}, BaF~\cite{Altunta, Bu, Parul1} molecules. Heavier members of this series, particularly BaF and RaF, have been studied for the nuclear anapole moment~\cite{Nayak, Kudashov, Isaev, Hao}, another parity-violating effect. The theoretical as well as experimental results of spectroscopic constants, valence properties and vibrational parameters for the ground\,- and excited states of AEMFs have also been reported by several research groups % at various levels of theories
~\cite{Isaev, Tohme, Jardali, Kork, Fowler, Bala, Meyer, Kosicki, Parul, Pelegrini, Bundgen, Machado, Ornellas, Gao, Rice, Childs, Sheridan, Bloch, Wall}. Nevertheless, there is a scope to carry out the calculations of $\mathcal{P}$ \& $\mathcal{T}$\,-odd interaction constants of these molecules consistently employing accurate many-body methods as there are only a very few calculations available in the literature. Nayak and Chaudhuri have reported the values of $W_d$ constant for the ground state of BaF system using the Dirac\,-\,Fock (DF) and the restricted active space configuration interaction (RASCI) methods~\cite{Nayak}, Kudashov \textit{et al.} have reported the \textit{ab initio} calculations of $W_d$ and $W_s$ constants for RaF molecule using Coupled cluster (CC) method~\cite{Kudashov}, Kozlov \textit{et al.} have done semi\,-\,empirical, and \textit{ab initio} calculations at self\,-\,consistent field (SCF) and restricted active space SCF (RASSCF) levels to report the $\mathcal{P}$ \& $\mathcal{T}$\,-odd interaction constants of BaF system~\cite{Kozlov, Kozlov1}, Isaev \textit{et al.} have computed the $W_d$ and $W_s$ results for BaF and RaF molecules using two\,-\,component zeroth order regular approximation (ZORA) together with generalized Hartree\,-\,Fock (HF) method~\cite{Isaev1}. Further, the CC calculations in Z-vector and expectation value approach have been performed by Sasmal \textit{et al.} to report the $\varepsilon_{eff}$ and $W_s$ constant for the RaF system~\cite{Sasmal}. Recently, Abe \textit{et al.} have performed the calculations of $\varepsilon_{eff}$ for AEMF (AE\,=\,Be, Mg, Ca, Sr, Ba) molecules using linearized CC (LCC) and finite\,-\,field CC (FFCC) methods~\cite{Abe_AEF}. There are no results available in the literature, however, for $W_s$ constants of the lower members of AEMFs.
%%%%%%%%%%%%%%%%%%%%%

In the present work, we have performed the calculations of $\mathcal{P}$ \& $\mathcal{T}$ -odd interaction constants in AEMFs using Kramers\,-\,restricted configuration interaction method restricted to single and double excitations (KRCISD) together with the quadruple zeta (QZ) quality basis sets. In Ref.~\cite{Bala}, we have applied this method to compute the valence properties: permanent dipole moments (PDMs) and dipole polarizabilities of AEMFs, and the atomic polarizabilities of AE and fluorine atoms. 
%%%%%%%%%%%%%%%%%%%%%

This paper is organized in three other sections: the following Section~\ref{section-2} discusses theory and method used for the calculations of symmetry violating constants,  followed by the detailed discussion on the computed results in Section~\ref{section-3} and the summary of the present work in the last section.
%-------------------------------------------%
\section{Theory and Method of Calculations}
\label{section-2}
%-------------------------------------------%
%%%%%%%%%%%%%%%%%%%%%%%%%%%%%%%%%%%%%
%%%%%%%%%%%%%%%%%%%%%%%%%%%%%%%%%%%%%%%%
\subsection{$\mathcal P$ \& $\mathcal T$\,-Odd Interaction Constant Relevant to eEDM} 
%%%%%%%%%%%%%%%%%%%%%%%%%%%%%%%%%%%%%%%%
The $\varepsilon_{eff}$ arises from the relativistic interactions of the eEDM with the electric fields created due to all other charged particles in a molecular system. 
The expectation value of the operator describing the interaction of eEDM in a molecular system is given by~\cite{Fleig, Fleig1, Das, Commins, Prasannaa},
\begin{eqnarray}{}
\Delta U\,=\,\bigg\langle\sum\limits_{j\,=\,1}^{n}\hat{H}_{EDM}(j)\bigg\rangle_{\Psi}\,=\,-\,d_e\,\bigg\langle\sum\limits_{j=1}^{n}\,\gamma_j^0\,\Sigma_j\,\varepsilon_j\bigg\rangle_{\Psi} \nonumber\\
\approx  \frac{2icd_e}{e\hbar}\bigg\langle\sum\limits_{j\,=\,1}^{n}\gamma_j^0\,\gamma_j^5\,p_j^2\bigg\rangle_{\Psi}
\end{eqnarray}
where $d_e$ is the electric dipole moment of an electron; $\gamma^0$ and $\gamma^5$ are the $4$\,-\,component Dirac matrices; $\varepsilon_j$ is the electric field at the position of $j^{th}$ electron; $p_j$ is the momentum operator; and $\Psi$ is the wavefunction determined from the many\,-\,body theory. Finally, the $\varepsilon_{eff}$ experienced by the unpaired electron in the molecular system is defined as,
\begin{eqnarray}{\label{Eeff}}
 \varepsilon_{eff}\,=\,W_d\,\Omega
\end{eqnarray}
where $W_d\,=\,({2ic}/{\Omega\,e\hbar})\langle\gamma^0\,\gamma^5\,p^2\rangle_{\Psi}$ is the $\mathcal P$ \& $\mathcal T$\,-odd interaction constant and $\Omega\,(=1/2)$ is the \textit{z}\,-\,component of the total angular momentum for the ground states of AEMFs. The intrinsic value of eEDM is calculated from the theoretically determined $\varepsilon_{eff}$ together with the experimentally measured energy shift ($\Delta U$) via equation $\Delta U\,\propto\,-\,d_e\,\varepsilon_{eff}$.
%%%%%%%%%%%%%%%%%%%%%%%%%%%%%%%%%%%%%%%%%%
\subsection{Scalar\,-\,Pseudoscalar Interaction Constant}
%%%%%%%%%%%%%%%%%%%%%%%%%%%%%%%%%%%%%%%%%%
The nucleon\,-\,electron interaction that arises due to the coupling between scalar\,-\,hadronic current and the pseudoscalar electronic current is known as scalar\,-\,pseudoscalar ($S$\,-\,$PS$) interaction. The $S$\,-\,$PS$ interaction Hamiltonian for any system is given by~\cite{Kudashov, Commins},
\begin{eqnarray}{\label{HS-PS}}
H_{S-PS}\,=\,\frac{i}{e}\frac{G_F}{\sqrt{2}}\,\sum\limits_{j\,=\,1}^{n}\sum_{A\,=\,1}^{N}\,k_{s,A}\,Z_A\,\gamma^0\,\gamma^5\,\rho_{A}(r_{Aj})
\end{eqnarray}
where G$_F$ (\,$=2.22249\,\times\,10^{-14}$\,$E_h\,a_0^{3}$) is the Fermi coupling constant~\cite{Kudashov, Sunaga}; $\rho_A$ is the nuclear charge density normalized to unity; the summation indices $j$ and $A$ 
run over the number of electrons and nuclei, respectively. The $k_{s,\,A}$ is a dimensionless electron\,-\,nucleus $S$\,-\,$PS$ coupling constant of an atom and it is defined as~\cite{Sunaga},
\begin{eqnarray}
k_{s,A}\,=\,k_{s,p}+\frac{N_A}{Z_A}\,k_{s,n}
\end{eqnarray}
where $Z_A$ and $N_A$ represent the number of protons and neutrons, respectively. $k_{s,p}$ ($k_{s,n}$) represent the $S$\,-\,$PS$ coupling constant of an electron and a proton (neutron). Further, the $\mathcal P$ \& $\mathcal T$\,-odd interaction constant ($W_s$) arising from the electron\,-\,nucleon $S$\,-\,$PS$ interaction can be evaluated as,
\begin{eqnarray}
W_s\,=\,\frac{1}{k_{s,\,A}\,\Omega}\,\langle H_{S-PS}\rangle_{\Psi}
\end{eqnarray}
%%%%%%%%%%%%%%%%%%%%%%

In order to compute the $\mathcal P$ \& $\mathcal T$\,-odd interaction constants, we have utilized CI method available in the KRCI module of DIRAC17 software suite~\cite{DIRAC}. 
%%%%%%%%%%%%%%%%%%%%%%%%%%%%%%%%%%%%%%%%%%%%%%%%%%%%%%%%%%%%%%%
%-------------------------------------------%
%%%%%%%%%%%-------Table-I--------%%%%%%%%%%%
%-------------------------------------------%
\begin{table}[ht!]
\begin{center}
\caption{\label{T_Basis} Details of the basis sets, in uncontracted form, used in this work.}
\begin{tabular}{llll}
\hline\hline
Atom & &&Basis \\
\hline
Be & &&cc\,-\,pVQZ: 12s,\,6p,\,3d,\,2f,\,1g\\
Mg & &&cc\,-\,pVQZ: 16s,\,12p,\,3d,\,2f,\,1g\\
Ca & &&dyall.v4z: 30s,\,20p,\,6d,\,5f,\,3g\\
Sr & &&dyall.v4z: 33s,\,25p,\,15d,\,4f,\,3g\\
Ba & &&dyall.v4z: 35s,\,30p,\,19d,\,4f,\,3g\\
Ra & &&dyall.v4z: 37s,\,34p,\,23d,\,15f,\,3g\\
F  & &&cc\,-\,pVQZ: 12s,\,6p,\,3d,\,2f,\,1g\\
\hline\hline
\end{tabular}
%\end{adjustwidth}
\begin{flushleft}
\end{flushleft}
\end{center}
\end{table}
%%%%%%%%%%%%%%%%%%%%%%
After generating the reference state using DF Hamiltonian, the generalized active space (GAS) technique is employed to perform KRCISD calculations. The Gaussian charge distribution for the nuclei is used in these calculations. Further, we have used the uncontracted correlation\,-\,consistent polarized valence quadruple zeta\,(cc\,-\,pVQZ)~\cite{Dunning} basis sets for low $Z$ elements: Be, F and Mg, and Dyall basis sets of similar quality (dyall.v4z)~\cite{Dyall_basis} for high $Z$ elements: Ca, Sr, Ba and Ra. These basis sets are significantly large, particularly when used in uncontracted form, as it can be seen from the explicit number of functions shown in Table~\ref{T_Basis}. 
%-------------------------------------------%
%%%%%%%%%%%-------Table-II--------%%%%%%%%%%%
%-------------------------------------------%
\begin{table*}[ht!]
\begin{ruledtabular}
\begin{center}
\caption{\label{T_GAS} Generalized active space model for the CI wavefunctions of AEMFs with $10E_h$ virtual cutoff energy.}
\begin{tabular}{l|lllllll}
Molecule & \multicolumn{5}{c}{Number of orbitals} & Number of & Number of\\
\cline{2-6}
& Frozen core & GAS1 & GAS2 & GAS3 & GAS3$^*$ & determinants & determinants$^*$\\
\hline
BeF & 2 & 4 & 1 & 80 & 132 & 410645 & 1116857\\
MgF& 6 & 4 & 1 & 84 & 146 & 452681& 1366127\\
CaF& 7 & 7 & 1 & 139 & 189 & 3789982& 7005482\\
SrF& 15 & 8 & 1 & 132 & 182 & 4463853& 8484303\\
BaF& 24 & 8 & 1 & 134 & 212 & 4600095& 11510973 \\
RaF& 40 &8 & 1 & 133 & 203 & 4531718 & 10554588\\
  \end{tabular}
\begin{flushleft}
$^*$For the case of aug-CV-QZ basis sets.
\end{flushleft}
\end{center}
\end{ruledtabular}
\end{table*}
%%%%%%%%%%%%%%%%%%

The DF orbitals having energy less than -$2\,E_h$ are considered as frozen core. The alkaline\,-\,earth atom is chosen as the coordinate origin of the corresponding diatomic molecule. In the GAS technique, active DF orbitals are divided into three subspaces: paired (GAS1), unpaired (GAS2), and virtual orbitals (GAS3). Further, cutoff energy of $10\,E_h$ is set uniformly for all molecules in order to truncate the higher virtual orbitals so as to make the computations manageable. The number of Slater determinants along with the number of active orbitals in different subspaces for all molecules are given in Table~\ref{T_GAS}.
%%%%%%%%%%%%%%%%%%%%%%%%%%

The values of equilibrium bond lengths used in the present work are: $1.359$ {\AA} for BeF~\cite{Kork}, $1.778$ {\AA} for MgF~\cite{Kork}, $2.015$ {\AA} for CaF~\cite{Kork}, $2.124$ {\AA} for SrF~\cite{Jardali}, $2.162$ {\AA} for BaF~\cite{Tohme}, and $2.244$ {\AA} for RaF~\cite{Isaev}. 

%-------------------------------------------%
\section{\label{section-3}Results and Discussion}
%-------------------------------------------%
%-------------------------------------------%
\subsection{$\mathcal P$ \& $\mathcal T$\,-Odd Interaction Constant Relevant to eEDM}
%-------------------------------------------%
The computed values of $\mathcal P$ \& $\mathcal T$\,-\,odd interaction constants, $W_d$, calculated at the KRCISD level of theory together with the available results in the literature are tabulated in Table~\ref{T_AEF_Wd}. The value of $W_d$ increases as we move from lighter to the heavier system due to increase in the difference between the $Z$ values of two atoms forming a diatomic molecule. Our results calculated at the KRCISD/QZ level compare well with the existing semi\,-\,empirical and \emph{ab initio} results reported in Refs.~\cite{Abe_AEF, Nayak, Isaev1, Kozlov, Kozlov1, Sasmal, Kudashov}.

We have also examined the effect of basis set augmentation and core\,-\,valence functions on $W_d$ by performing additional calculations using augmented\,-\,pCVQZ basis sets for lighter atoms and augmented\,-\,dyall.cv4z basis sets for heavy atoms. We referred these basis sets further as ``aug-CV-QZ". It can be seen from the last column of Table~\ref{T_GAS} that the number of Slater determinants increases significantly with the aug-CV-QZ basis sets. We have observed that the effect of adding extra functions to the QZ basis sets on the values of $W_d$ are about 1\% to all AEMFs, except BeF. The computational cost, on the other hand, increases considerably for the aug-CV-QZ calculations. To quote in this context, the amount of RAM required for the calculation of $W_d$ for RaF, at KRCISD/aug-CV-QZ level is about 500\% times larger than that required for the KRCISD/QZ level calculation, even when the number of filled active orbitals and the virtual energy cutoff are kept intact. 

%%%%%%%%%%%%%%%%%%%%%%%%%%%%%%%%%%%%%%
%-------------------------------------------%
%%%%%%%%%%%-------Table-III--------%%%%%%%%%%%
%-------------------------------------------%
\begin{table}[ht!]
\begin{center}
\caption{\label{T_AEF_Wd}$\mathcal P$, $\mathcal T$\,-odd interaction constant, $W_d$ (in Hz/e-cm) for the ground state of alkaline\,-\,earth monofluorides calculated at KRCISD level of theory, compared with the available calculations in the literature.}
\begin{tabular}{ccc}
\hline \hline
Molecule & {$\left|W_{d}\right |$ ($\times\,10^{25}$)} & Ref.\\
\hline
BeF & 0.00021$^*$ & This work\\
    & 0.00027$^{**}$ & This work \\
    & 0.00024$^{a\dag}$  & \cite{Abe_AEF}\\
MgF & 0.00269$^*$ & This work\\
    & 0.00272$^{**}$ & This work\\ 
    &  0.00339$^{a\dag}$ & \cite{Abe_AEF}\\
CaF & 0.01161$^*$ &This work\\
    & 0.01165$^{**}$ & This work\\ 
    &  0.01354$^{a\dag}$ &\cite{Abe_AEF}\\
SrF & 0.08979$^*$ & This work\\
    & 0.08964$^{**}$ & This work\\  
    &  0.10446$^{a\dag}$ & \cite{Abe_AEF}\\
BaF & 0.27426$^*$ & This work\\
    & 0.27321$^{**}$ & This work\\ 
    & 0.293$^b$, 0.352$^c$ & \cite{Nayak}\\
    & 0.31240$^{a\dag}$      & \cite {Abe_AEF}\\    
    & 0.26$^{d\dag}$ & \cite{Isaev1}\\
    & 0.230$^e$, 0.224$^f$, 0.375$^g$, 0.364$^h$ & \cite{Kozlov} \\
    & 0.35$^i$ 0.41$^i$& \cite{Kozlov1}\\
RaF & 2.34303$^*$ &This work\\
    & 2.33571$^{**}$ & This work\\
    & 2.40$^j$, 2.25$^k$, 2.65$^l$, 2.36$^m$,& \cite{Kudashov}\\
    &2.33$^n$, 2.30$^o$, 2.56$^p$ &\\
    & 2.20$^{d\dag}$    & \cite{Isaev1}\\
    & 2.54$^{q\dag}$, 2.55$^{r\dag}$&\cite{Sasmal}\\
  \hline \hline
   \end{tabular}
\begin{flushleft}
$^*$ These results are calculated using QZ quality basis sets.\\
$^{**}$ These results are calculated using aug-CV-QZ quality basis sets.\\ 
$^a$FFCCSD, $^b$DF, $^c$RASCI,
$^d$Using 2\,-\,component GHF\,-\,ZORA value of $W_S$, $^e$SCF,
$^f$RASSCF, $^g$SCF\,-\,EO, $^h$RASSCF\,-\,EO, $^i$semi\,-\,empirical calculations,
$^j$SODCI, $^k$FS\,-\,RCCS, $^l$FS\,-\,RCCSD, $^m$CCSD, $^n$CCSD(T), $^o$CCSD$_{enlarged}$, $^p$CC$_{final}$, $^q$Z-vector method in CC approach, $^r$Expectation value in CC approach.\\
$^\dag$These results are computed using Eq.~(\ref{Eeff}) with the values of $\varepsilon_{eff}$ taken from the corresponding References.
\end{flushleft}
\end{center}
\end{table} 
%%%%%%%%%%%%%%%%%%%%%%%%%%%%%%%%
Kozlov~\emph{et al.}~\cite{Kozlov1} have reported the semi\,-\,empirical values of $W_d$ constant for BaF molecule based on two different experimental values of hyperfine structure constants~\cite{Knight, Ryzlewicz} to be $0.35\,(\times\,10^{25}$\,Hz/e-cm) and $0.41\,(\times\,10^{25}$\,Hz/e-cm). Our \emph{ab initio} result of  $W_d\,=\,0.27426\,(\times\,10^{25}$\,Hz/e-cm) differs from those reported in their work by $0.07574$\,($\times\,10^{25}$\,Hz/e-cm) and $0.13574$\,($\times\,10^{25}$\,Hz/e-cm). Later, in Ref.~\cite{Kozlov}, Kozlov~\emph{et al.} have performed RASSCF calculations by considering $11$ electrons in three RASs: RAS1\,(2,0,0,0), RAS2\,(2,1,1,0) and RAS3\,(6,4,4,2). Further, they have used effective operators (EOs) to include the core\,-\,polarization effects. Their final value of $W_d$ $=\,0.364\,(\times\,10^{25}$\,Hz/e-cm) at RASSCF\,-\,EO level, which is larger from that computed in our work at KRCISD/QZ level by $0.08974\,(\times\,10^{25}$\,Hz/e-cm).
%%%%%%%%%%%%%%%%%%%%%%%%%%%%%%%%

Nayak and Chaudhuri~\cite{Nayak} have utilized the RASCI method together with the uncontracted Gaussian basis sets (27s\,27p\,12d\,8f) for Ba and (15s\,10p) for F to compute the $W_d$ constant of BaF. Further, they have considered $17$ electrons in 76 active orbitals. We have, on the other hand, considered the same number of electrons in $143$ active orbitals, which is very large in comparison to that included in Ref.~\cite{Nayak}. Our computed result using QZ basis sets differs from their result by $0.07774$\,($\times\,10^{25}$\,Hz/e-cm) at the similar level of correlation.
%%%%%%%%%%%%%%%%%%

Isaev and Berger~\cite{Isaev1} have obtained the $\varepsilon_{eff}$ for BaF and RaF numerically by using relationship between matrix elements of $\mathcal{P}$ \& $\mathcal{T}$\,-odd and $\mathcal{P}$\,-odd operators. Our value of $W_d$ parameter is larger by $5.5$\%\,($6.5$\%) for BaF\,(RaF) from that computed using the results of $\varepsilon_{eff}$ reported in their work via Eq.~(\ref{Eeff}). 
%%%%%%%%%%%%%%%%%%

Kudashov \emph{et al.}~\cite{Kudashov} have performed spin\,-\,orbit direct CI (SODCI) calculations by considering $19$ electrons explicitly to report the $W_d$ constant for RaF and our result varies by $2.4$\% from that estimated in their work. In the same work, those authors have also reported the $W_d$ constant using relativistic $2$\,-\,component Fock\,-\,Space CC method by considering single and double excitations (FS-CCSD). Our value of $W_d$ at KRCISD/QZ level differs by 11.9\% from that reported at FS-CCSD level in Ref.\cite{Kudashov}. Further, they have corrected FS-CCSD result by including the contributions due to triple excitations as well as basis set enlargement. Their final value (=\,$2.56$\,($\times\,10^{25}$\,Hz/e-cm)) of $W_d$ including these corrections is larger by $8.5$\% from our result (=\,$2.34303$\,($\times\,10^{25}$\,Hz/e-cm)) using the relativistic Hamiltonian at the KRCISD/QZ level.
%%%%%%%%%%%%%%%%%%

Sasmal~\emph{et al.}~\cite{Sasmal} have applied the expectation value and Z-vector approach in CC framework to compute the $\varepsilon_{eff}$ of RaF molecule. Further, they have considered single and double excitations together with the dyall.cv3z and dyall.cv4z basis sets for Ra and cc-pCVTZ and cc-pCVQZ basis sets for F atom. Our value of $W_d$ at KRCISD/QZ is smaller by 7.8\% and 8.1\% than their results using Z-vector method and expectation value approach, respectively, at the similar level of basis sets.
%%%%%%%%%%%%%%%%%

Abe~\emph{et al.}~\cite{Abe_AEF} have reported the values of $\varepsilon_{eff}$ for the ground states of AEMFs: BeF, MgF, CaF, SrF and RaF using FFCCSD method. The authors have performed all\,-\,electron calculations using cc-pVNZ (N\,=\,D,\,T,\,Q) basis sets for Be, Mg, Ca and F, and a combination of Dyall and Sapporro basis sets for heavier elements: Sr and Ra. However, we have compared, in Table-\ref{T_AEF_Wd}, our results with theirs computed at the QZ level only, for a fair comparison. The maximum difference between our results and those reported in their work is $0.03814$\,($\times\,10^{25}$\,Hz/e-cm) for BaF.
%%%%%%%%%%%%%%%%%%
%-----------------------------------------%
\subsection{Scalar\,-\,Pseudoscalar Interaction Constant}
%-------------------------------------------%
The calculated $S$\,-\,$PS$ $\mathcal P$ \& $\mathcal T$\,-odd interaction constants ($W_s$) for the ground states of AEMFs at 
%%%%%%%%%%%%%%%%%%
%-------------------------------------------%
%%%%%%%%%%%-------Table-IV--------%%%%%%%%%%%
%-------------------------------------------%
\begin{table}[]
\begin{center}
\caption{\label{T_AEF_Ws}$S$\,-\,$PS$ constant, $W_s$ (in KHz) for alkaline\,-\,earth atoms in alkaline\,-\,earth monofluorides calculated at KRCISD level of theory, compared with the available results in the literature.}
\begin{tabular}{cccc}
\hline \hline
Molecule & Atom& $W_{s}$    & Ref.\\
  \hline 
  BeF      & Be&  0.00132$^*$  & This work \\
           &   & 0.00142$^{**}$  & This work \\
           & F  & 0.00301$^*$       & This work\\
           &    & 0.00317$^{**}$  & This work \\
  MgF      & Mg&0.04138$^*$   & This work\\
           &   & 0.04116 $^{**}$  & This work \\
           & F   &   0.00408$^*$      & This work\\ 
           &     & 0.00428$^{**}$  & This work \\  
  CaF      & Ca&0.17751$^*$  & This work\\
           &   & 0.17699$^{**}$  & This work \\
           & F   & 0.00131$^*$  & This work\\
           &     & 0.00137$^{**}$  & This work \\
  SrF      & Sr&1.70329$^*$ & This work\\
           &   &1.70028$^{**}$  & This work \\
           & F  &0.00098$^*$  &       This work\\
           &    &0.00102$^{**}$  & This work \\ 
  BaF      & Ba&7.28604$^*$ & This work\\
           &   &7.25918$^{**}$  & This work \\ 
           &&     11$^a$, 13$^a$      &\cite{Kozlov1}\\
          & &     6.1$^{b}$, 5.9$^{c}$           & \cite{Kozlov}\\
          &&8.5$^d$&\cite{Isaev1}\\
           & F  &0.00025$^*$& This work\\
           &    &0.00025$^{**}$  & This work \\
  RaF      & Ra&130.52357$^*$ & This work\\
           &   &130.12190$^{**}$  & This work \\
         &  &     131$^e$, 122$^f$, 144$^g$, 128$^h$,&\cite{Kudashov}\\ 
        &&  127$^i$, 125$^j$, 139$^k$   \\
        && 150$^d$&\cite{Isaev1}\\
        && 141.2$^l$, 142$^m$&\cite{Sasmal}\\
           & F  &0.00066$^*$ & This work\\ 
           &    & 0.00069$^{**}$  & This work \\
           \hline \hline
\end{tabular}
\begin{flushleft}
$^*$ These results are calculated using QZ quality basis sets.\\
$^{**}$ These results are calculated using aug-CV-QZ quality basis sets.\\ 
$^a$Semi\,-\,empirical calculations, $^{b}$SCF, $^{c}$RASSCF, 
$^d$two\,-\,component ZORA generalized Hartree\,-\,Fock (GHF), $^e$SODCI, $^f$FS\,-\,RCCS, $^g$FS\,-\,RCCSD, $^h$CCSD, $^i$CCSD(T), $^j$CCSD$_{enlarged}$, $^k$CC$_{final}$, $^l$Z-vector method in CC approach, $^m$Expectation value in CC approach.
\end{flushleft}
\end{center}
\end{table}
%%%%%%%%%%%%%%%%%
the KRCISD level of correlation along with the available results in the literature are given in Table~\ref{T_AEF_Ws}. It is clear from Eq.~(\ref{HS-PS}) that the matrix element of $H_{S\,-\,PS}$ varies roughly as $\displaystyle A\,Z^{2}\,\approx\,Z^3$~\cite{Commins} and hence, heavy polar molecules are preferred for the study of symmetry violating effects.  As can be seen from Table~\ref{T_AEF_Ws}, the contribution of the lighter atom, \emph{viz.} fluorine, in a diatomic molecule is negligibly small. On the other hand, as one moves from BeF to RaF, the contribution of AE atom to $W_s$ increases. The value of $W_s$ calculated using aug-CV-QZ basis sets for Be is larger by $7.6\%$ from that computed using QZ basis sets. However, for all other AE atoms, the inclusion of extra functions to the QZ basis sets lower the values of $W_s$ by about $1$\% or less. The difference between the values of $W_s$ using aug-CV-QZ and QZ basis sets for fluorine is 5.3\% in BeF, 4.9\% in MgF, 4.6\% in CaF, 4.1\% in SrF and 4.5\% in RaF, whereas for the case of BaF, it does not change atleast up to the accuracy reported in our work. 

There are no calculations available in the literature to compare our results for the first four members of the series. However,  our value for BaF lies in between the results reported in Refs.~\cite{Isaev1, Kozlov, Kozlov1}. Our result of $130.5236$\,KHz at KRCISD/QZ level for RaF is very close to $131$\,KHz reported in Ref.~\cite{Kudashov} at SODCI level. The final CC value that includes triple contributions as well as large basis set effects estimated in Ref.~\cite{Kudashov}, is $6.1$\% larger than our result. However, the calculated value of $W_s$ using two\,-\,component ZORA generalized HF method in Ref.~\cite{Isaev1} is larger by 13\% than our result. Our result of $W_s$ for RaF system is smaller by $7.6$\% and $8.1$\% than computed using Z-vector method and as an expectation value, respectively in Ref.~\cite{Sasmal}. 
%------------------------------------------%
\section{\label{section-4}Summary}
%-------------------------------------------%
%%%%%%%%%%%%%%%%%%%%%%
In summary, we have performed relativistic calculations of $\mathcal{P}$ \& $\mathcal{T}$\,-odd interaction constants: $W_d$ and $W_s$ in AEMFs using KRCISD method in conjunction with the quadruple zeta quality basis sets. Further, the effect of adding diffuse as well as core\,-\,valence functions to the QZ basis sets on $\mathcal{P}$ \& $\mathcal{T}$\,-odd interaction constants are studied and we have observed that the results of $W_d$ of AEMFs will not be affected by more than $~1$\% while the results of $W_s$ of AE atoms, the change is also about $1$\% or less, with an exception of BeF. Our results at KRCISD/QZ level of the theory show reasonably good agreement with most of the existing \textit{ab initio} calculations in the literature. We thus believe that the results reported in this work would be useful for the future theoretical and experimental studies relevant for the search of electric dipole moment of an electron in these molecules.

\begin{center}
{\bf {ACKNOWLEDGMENTS}}
\end{center}
A major part of the calculations reported in this work were performed on the computing facility available in the Department of Physics at IIT Roorkee, India. This research was supported by Department of Science and Technology, Inspire and FIST division, India (Grant No. SR/FST/PS1-148/2009(C)). 
\end{document}